\documentclass[3p,twocolumn]{elsarticle}
\usepackage{graphicx}
\usepackage{amsmath}
\usepackage{amssymb}

\begin{document}
\title{Universal features of surname distribution in a subsample of a growing population}
\author{Yosef E.  Maruvka}
\ead{yosi.maruvka@gmail.com}
\author{Nadav M. Shnerb}
\ead{shnerbn@mail.biu.ac.il}
 \author{David A. Kessler\corref{cor1}}
 \ead{kessler@dave.ph.biu.ac.il}
 \cortext[cor1]{Corresponding author, (tel) +972-3-531-8177, (fax) +972-3-738-4054}
\address{Department of Physics, Bar-Ilan University,
 Ramat-Gan 52900 Israel }

\journal{Journal of Theoretical Biology}

\begin{abstract}
We examine the problem of family size statistics (the number of
individuals carrying the same surname, or the same DNA sequence) in
a given size subsample of an exponentially growing population. We
approach the problem from two directions.  In the first, we
construct the family size distribution for the subsample from the
stable distribution for the full population.  This latter
distribution is calculated for an arbitrary growth process in the
limit of slow growth, and is seen to depend only on  the average and
variance of the number of children per individual, as well as the
mutation rate. The distribution for the subsample is shifted left
with respect to the original distribution, tending to eliminate the
part of the original distribution reflecting the small families, and
thus increasing the mean family size.  From the subsample
distribution, various bulk quantities such as the average family
size and the percentage of singleton families are calculated.  In
the second approach, we study the past time development of these bulk
quantities, deriving the statistics of the genealogical tree of the
subsample.  This approach reproduces that of the first when the
current statistics of the subsample  is considered.  The surname distribution from th e 2000 U.S. Census is examined in light of these findings, and found to misrepresent the population growth rate by a factor of 1000.
\end{abstract}


\maketitle

\def\baselinestretch{1}

\section{Introduction}

There is  a long history of work in the social sciences on family
size distributions. The classic founding work in this field is that
of Galton and Watson (GW) ~\citep{Gal}  who tried to explain the
decline of the British great families, as indicated by data on
surname abundance.  Rejecting previous explanations based on
"fitness", e.g., that the rise of physical comfort is followed by
fertility decline, they assumed that the phenomenon is purely
statistical.  The affiliation of an individual with certain family,
expressed in his/her surname, was assumed by GW to be  a neutral
property. This feature is inherited  to the next generation
according to a well defined rule (all offsprings take the surname of
their father) and is subject to the stochasticity that characterizes
birth-death processes. Assuming a well-mixed population, GW claimed
that all surnames undergo extinction in the long run. In fact, their
conclusions were correct only for an equilibrium population, whereas
for a growing population,  their equations exhibit a second
nontrivial solution which was found by Steffenson ~\citep{Steffensen30,Steffensen33}
and exploited by Lotka ~\citep{Lotka, Lotka2} using U.S. census data to deduce the offspring distribution.
Subsequently the impact of  surname changes (``mutations") was
considered by Manrubia and Zanette (MZ)~\citep{MZ}. All in all, the
surname in a society undergoes a birth-death-mutation process, and
the current surname abundance distribution reflects the demographic
(birth-death ratio) and social (mutation rate) characteristics of
the population. MZ also presented data for the distribution of
surname in the populations in Argentina, Berlin, and five cities in
Japan, where the statistics were obtained from  phonebooks.  The
data exhibited the predicted $1/n^2$ behavior at large $n$ for the
probability of $n$ appearances of a surname.  MZ then attempted to
use the deviations for smaller $n$ to deduce the growth rate of the
population.

As already pointed out in \citet{Der}, the importance of the clan
statistics for a population that undergoes a birth-death-mutation
process goes far beyond its applicability to surname dynamics. Any
neutral genetic feature associated with a sequence that appears on
certain loci and is subject to mutations undergoes exactly the
same process, thus the results for surnames reflect also the amount
of genetic polymorphism in the population. Another neutral process
of the same  kind was suggested by Hubbell \citep{hub2,hub1,bell} and
Bell \citep{bell} as the underlying mechanism that yields the
observed species abundance distribution. This heretical idea opposes
the traditional ``niche" theories that seeks to explain species abundance
ratios in terms of interspecies interaction and fitness, and ignited
an enormous  contentious debate on that subject \citep{Mc}.
The argumentation of both sides is based on the species abundance
statistics, as gathered in large-scale censuses \citep{1} and the
very same problem arises: what statistics should one expects in case
of a growing or shrinking population which is subject to neutral
mutation?

Before trying to compare the observed statistics with some
theoretical predictions (e.g., in order to recover demographic
parameters from the abundance ratio) one should address two crucial
issues. The first is \emph{universality}: to what extent should one
expect the results to be independent of the "microscopic" features
of the process? Fig. 1 shows the family size statistics (Pareto plot) obtained
from numerical simulations of two populations with the same
demographic characteristics. The dynamics  assumes nonoverlapping
generations, where the average number of offspring per individual is
$\lambda > 1$ and the chance of an offspring to mutate (i.e., to
differ from its originator and to start a new clan) is $\mu$. Both
populations have the same values for $\lambda$ and $\mu$, but they
differ microscopically: in one case the chance for an individual to
produce $n$ offsprings obeys the Poisson distribution with  average
$\lambda$, in the other case it satisfies the geometrical distribution with
the same average. As one can clearly see, the tails of these
distribution coincides, but  the  bulk abundance statistic  is
\emph{different}; this implies that the theory of abundance ratio
has no use for any practical purpose unless one knows the very
fine details of the demographic properties of the population throughout
history, an inconceivable task in almost all circumstances. A comparison
between experimental  data and theoretical predictions is possible
if, and only if, one can show that there is a universal regime in
which the statistics is independent of the microscopic details; this
is one of the aims of this paper.

\begin{figure}
\center{\includegraphics[width=0.5\textwidth]{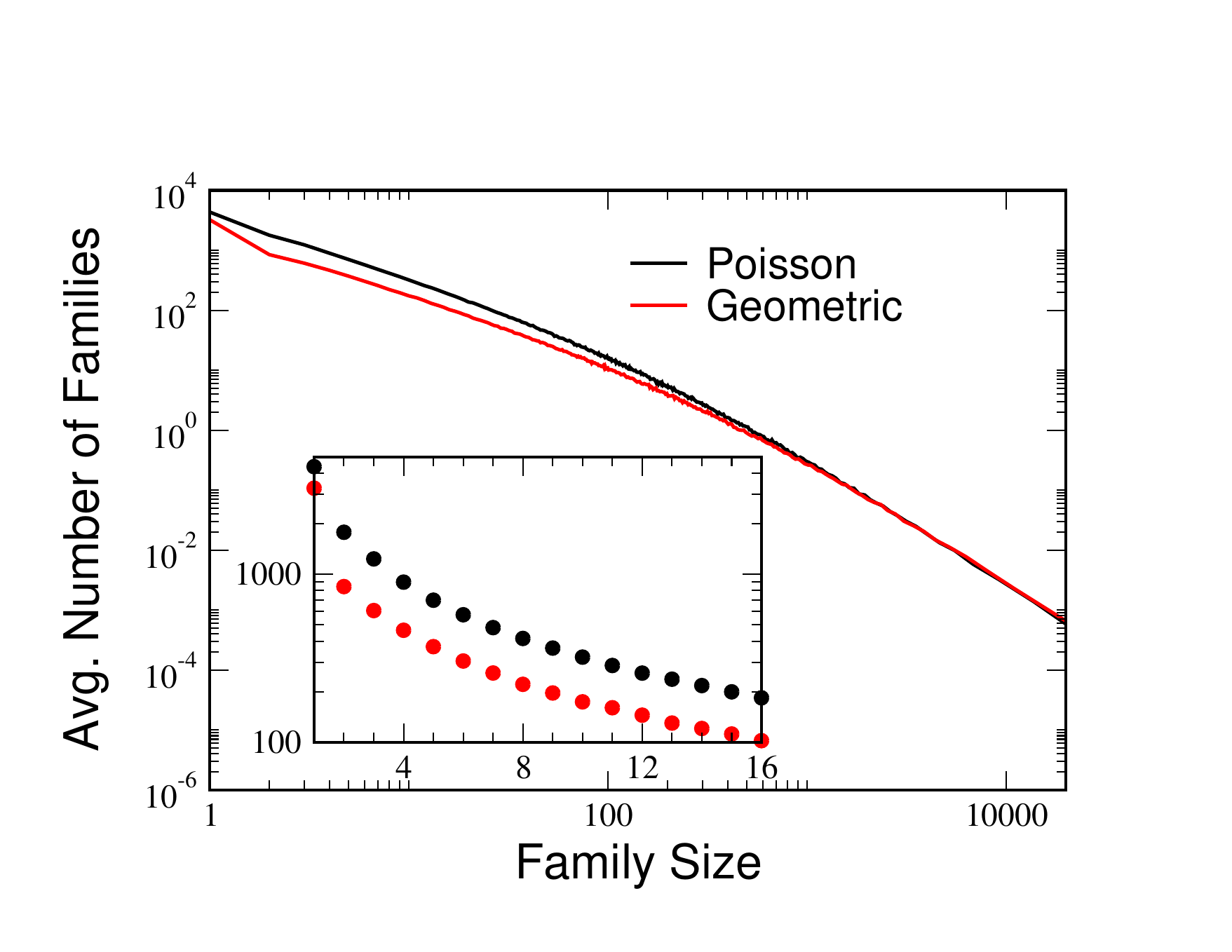}}
\caption{Family size distribution for a population of $4\cdot10^6$,
for the parameters $\lambda=1.005$, $\mu=5\cdot10^{-4}$, for a
Poisson and for a geometrical distribution of offspring.  The data
was averaged over 100 runs and binned into bins which contained a
minimum of 1000 families over the 100 runs. One sees that the large
families are distributed in both cases as a power law, as in the MZ
model.  The power-law cuts off at small family sizes, below sizes of
roughly 1000, at which point the two distributions diverge. Inset:  A blowup of the
figure for small family sizes, highlighting the difference between the two distributions. }
\label{fsdistfig}
\end{figure}

The second issue that should be addressed is the effect of
\emph{sampling}. In all cases considered above - surnames, genetic
polymorphism, species abundance - the raw data is made of
individuals sampled from the whole population together with their
affiliation with certain surname or certain species. It is difficult to
perform a complete census, given that typically one does not have
access to the entire population.  Thus for example, MZ only looked
at the surnames beginning with ``A" in the Berlin phone book. Even for the US
census data, one has access to (almost) the entire population only under the assumption that the US is demographically isolated, which it is clearly not.  In the
application we have in mind, that of looking at genomic data to
measure historic growth rates of the population, one has such data
for an extremely small sample of the entire human population. 

\begin{figure}
\center{\includegraphics[width=0.5\textwidth]{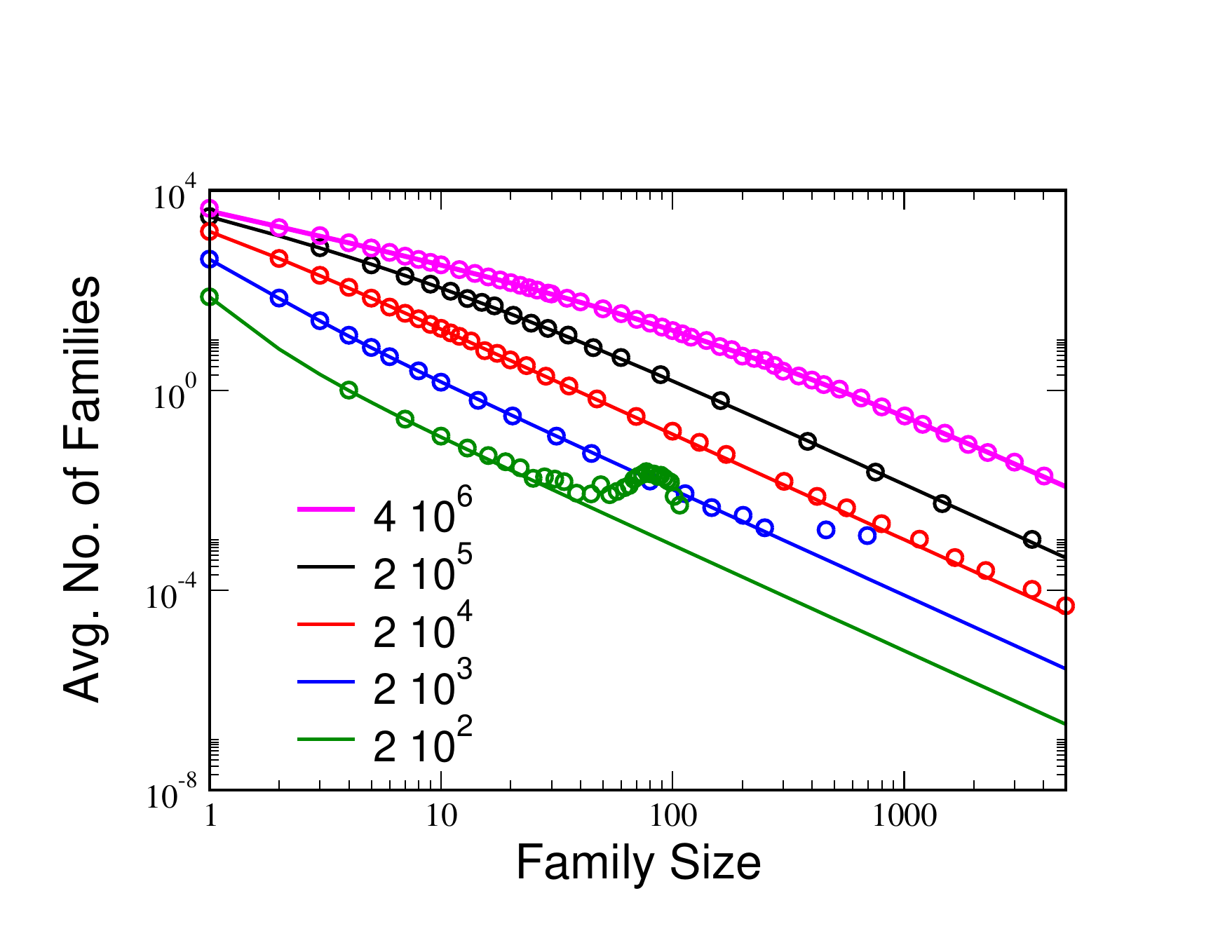}}
\caption{Avg. number of families of a given size, for the full population of $4\cdot 10^6$, and subsamples of size
$2\cdot 10^5$, $2\cdot 10^4$, $2\cdot 10^3$ and $200$.  The growth rate is $\gamma=0.005$, and the mutation rate is $\mu=5\cdot 10^{-4}$, and the child distribution is Poisson.  The circles represent averages over 100 iterations.  The lines are the theory for the full population, Eq. (\ref{fulln}) and for a ``red" subsample, Eq. (\ref{subdistn}). The deviations from the power law for the largest $m$'s seen in the $N_o=200$ and $2000$ data are due to the fact that the largest family does not obey the stable distribution, but rather reflects the single individual initial conditions chosen~\citep{MZ}.} \label{fig2}
\end{figure}

What is the effect of incomplete sampling? In Fig. \ref{fig2} one
can see the characteristic features of the family statistics
obtained in the two regimes:  strong and weak sampling. One can see
that the full statistics is characterized by a "shoulder" in the
small families region, followed by a power-law decay for large
families. If the sampling is strong the distribution is shifted
quite rigidly to the left, while the case of weak sampling is
characterized by a  peak for the singletons (families with
only one member) followed by a power-law. Our second aim here is to
clarify  effect in both regimes.

In the following, we analyze the problem from two different angles.  The first is
centered on the stable distribution for the entire population.  This
distribution can be calculated from a Fokker-Planck equation, akin
to that written down by MZ.  We show this Fokker-Planck equation  is
in fact {\emph{universally}} valid in the limit of small growth
rate, for an arbitrary distribution of children produced by an
individual, with the coefficients depending only on the average and
variance of the children distribution, together with the mutation
rate.  From this we can calculate the distribution for a
given sized subsample of the population in terms of a hypergeometric function.  
We then endeavor to assimilate the meaning of this result, focussing on the strong and weak sampling limits, exhibiting simpler formulae for the average family
size and number of singleton families
We also show that the
large-family power-law asymptotics is left unchanged by the
sampling.  

The second approach is based on looking at the behavior of the
genealogical tree of the selected sample.  We calculate the size of
the tree as a function of time, as well as the number of families
and singletons, all in the limit of small mutation rate.  These
results are seen to reproduce those of the previous approach for the
current statistics of the selected sample.

The plan of the paper is as follows.  In Section \ref{secmodel}, we
describe our model, explain the notation used along this work and
highlight our main results.  In Section III, we present our
derivation of the family size distribution for the subsample, and
calculate the average family size and number of singleton families.
In Section IV, we present our second approach. Finally, in Section V
we examine the surname distribution taken from  the U.S. census data in light of our findings.  We then summarize and present some concluding remarks.

\section{ Model, simulation technique and main results \label{secmodel}}

 Our basic model is that of a growing population with
nonoverlapping generations, as in the original Galton-Watson work.
Every member of the population simultaneously gives birth to a
random number of children, drawn from a given distribution with mean
$\lambda$, and is then removed.  The children are all reckoned to
belong to the same family as the parent, unless they undergo a
mutation at birth, which occurs with probability $\mu$.  The mutated
child is considered to start a new family.  We start the population
with one individual, repeating the experiment until the population
survives the initial stages and achieves the desired size.    In
principle, we could track the genealogy of every individual.  In
practice, for efficiency's sake, we track only the genealogy of
families, which is sufficient to determine the family identification
of every individual.  Thus, it is sufficient to draw the number of
children of each family.  In our simulations, we mostly employ a
Poisson distribution for the number of children, occasionally
comparing to the case of a geometrical distribution.  In the former
case, the distribution of children of a given family of size $n$ is
again Poisson, with mean $\lambda$, whereas for the geometrical
distribution, it is a Pascal (or negative binomial) distribution.

As a technical point, we will be interested in Section V in the
genealogical tree of subsamples of the population, so that we can
track the time development of the statistics of this tree.  We can
do this retrospectively for the Poisson case, simply picking
ancestors for each individual among the set of individuals in the
family in the previous generation which gave rise to this individual
(which is the same family, barring mutations). This is done to avoid
having to store the genealogies of individuals, which are of course
more voluminous than those of families.

\textbf{Glossary:} The growth rate $\gamma \equiv \lambda -1$ reflects
the deviation of the process from demographic equilibrium. In
general, as discussed above, the distribution function depends on
the details of $P(m)$, the chance of an individual to have $m$
children in the next generation.  It turns out, however,  that in
the universal regime the family statistics depends only on three
parameters: $\gamma$ (or equivalently $\lambda$), which reflects the
average number of offspring per individual, $\sigma$, the standard
deviation of the offspring distribution defined as
\begin{equation}
\sigma^2\equiv \sum_m m^2 P(m) - \lambda^2 = \textrm{Var}(m),
\end{equation}
 and $\mu$, the mutation
rate. For convenience, we define
\begin{equation}
\nu \equiv \frac{\mu}{\gamma-\mu}
\end{equation}
as this combination appears often.

The number of families with $m$ members is defined as
$n_m$. These definition implies  that the
sum of  $m n_m$  over all $m$'s yields the overall current size of the
population, $N_o$.  Except for $m$'s of order unity, one may consider the size of the
family as a continuous variable, thus replacing $m$ by $x$ and $n_m$
by $n(x)$. When the sampling is incomplete we tag the sampled
individuals as "red", defining $n^R_m$ as the abundance of families
with $m$ individuals in the red (sampled) population [When dealing
with the whole genealogy we define a "red" subgenealogy  consisting of all
the individuals that have at least one descendent in the sampled
population]. Sampling introduces a new parameter to the problem,
$R_o$, the number of sampled (``red") individuals. It turns out that there is a ``critical" sample size which distinguishes between weak and strong sampling:
\begin{equation}
R_c \equiv \frac{2N_o \gamma}{\sigma^2 (1+\nu)}
\end{equation}
We then measure sampling strength, $s$, through
\begin{equation}
s\equiv \frac{R_o}{R_c} = \frac{R_o\sigma^2 (1+\nu)}{2N_o\gamma}.
\label{sdef}
\end{equation}

Our main results are:

\begin{enumerate}
 \item  In the large $m$ limit $n_m$ decays like a
 power-law,  $n_m \sim m^{-\beta}$ where
\begin{equation}
\beta = \frac{\ln \lambda^2 (1-\mu)}{\ln \lambda(1-\mu)}
\label{betaeqnn}
\end{equation}
This law is semi-universal, in that it is independent of the details of $P(m)$. It however depends on the assumption of nonoverlapping generations, and therefore differs from the power law found by MZ.  It does however reduce to the  MZ result, $\beta = 2
+ \nu$, in the  slow growth, small mutation limit $\gamma \sim \mu \ll 1$.

\item  The whole distribution (except for the very smallest $m$'s)  becomes
universal if both $\mu$ and $\gamma$ are small. In that case $n(x)$
satisfy the Fokker-Planck equation:
\begin{equation} 
\frac{\partial n}{\partial t} = -(\gamma - \mu)
\frac{\partial}{\partial x} (xn) + \frac{\sigma^2}{2}
\frac{\partial^2}{\partial x^2} (xn). 
\label{FPn}
\end{equation}
A similar equation has been obtained by MZ for their particular model;
here we show that it is a universal limit of the process for small
rates, and also reveal its dependence on $\sigma$.

\item  Solving for the steady state distribution of (\ref{FPn}), the
 abundance distribution function is:
\begin{equation}
n(x) =  \frac{\nu R_c }{x}\Gamma\left(2+\nu\right)\,
U\left(1+\nu,0,\frac{2\gamma}{\sigma^2(1+\nu)}x\right) \label{fulln}
\end{equation}
where $U$ is the Kummer function \citep{AbSt}.   Thus, the abundance distribution for different microscopic processes with the same $\gamma$ and $\mu$ are related by a rescaling of the family size $m$ and the abundance $n$,
$n\sigma^4$ being a universal function of $m/\sigma^2$ (since $R_c \propto 1/\sigma^2$).  We see this in Fig. \ref{scaledfs}.

\begin{figure}
\center{\includegraphics[width=0.5\textwidth]{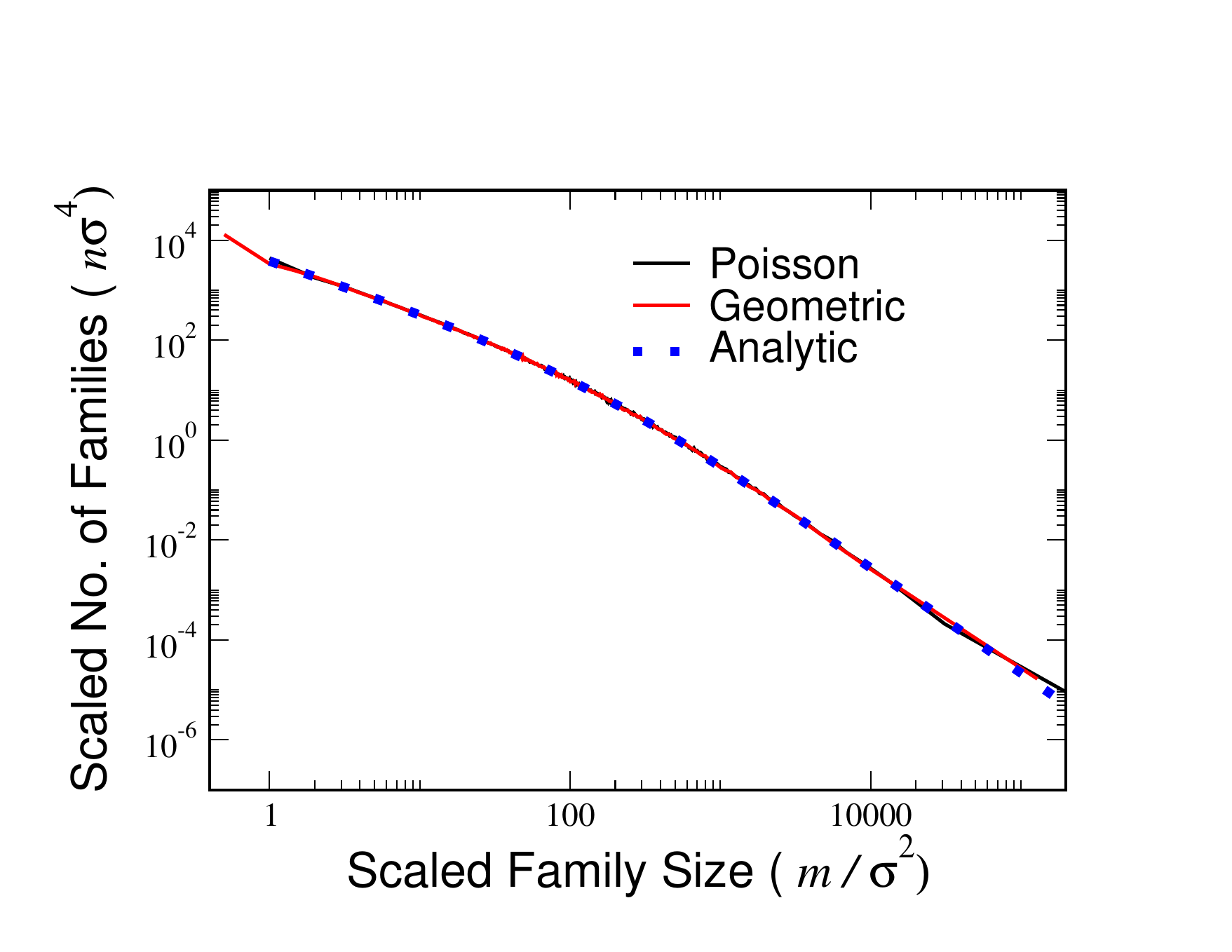}}
\caption{Scaled average number of families $n\sigma^4$ as a function
of scaled family size $m/\sigma^2$ for the Poisson and geometric
offspring distributions.  The data is the same as in Fig.
\ref{fsdistfig}.  Also shown is the analytic prediction, Eq.
(\ref{fulln}).} \label{scaledfs}
\end{figure}

\item  For the sampled (``red") population, $n^R_m$ is given by the
monstrous expression:
\begin{eqnarray}
\hspace*{-.25in}n^R_m \approx  \nu R_c  \textrm{B}\left(2+\nu,m\right)s^m\, {}_2F_1\left(m,m+1;m+2+\nu; 1 - s\right)\nonumber\\
&\ & \label{subdistn}
\end{eqnarray}
where $\textrm{B}(a,b)\equiv \Gamma(a)\Gamma(b)/\Gamma(a+b)$ is the
Beta function and ${}_2F_1$ is the hypergeometric function~\citep{AbSt}, and $s$ is the sampling strength introduced above. To digest this, we focus on two limits, that of strong and weak sampling.  In the
strong sampling limit, $\gamma N_o \ll R_o \ll N_o$, so that $s \gg 1$,
\begin{equation}
n_m^R \approx \nu R_c \frac{\Gamma(2+\nu)}{m}U(1+\nu,0,m/s)
\end{equation}
Thus, since $s$ is proportional to $R_o$, when varying $R_o$, $R_o n_m^R$ is a function only of $m/R_o$, and the dependence of
$R_o$ just amounts to rescalings of $m$ and $n_m^R$.  This implies that in this
limit the breakdown of the asymptotic power-law occurs at $m$'s of
order $N_o\gamma/R_o$, and in general sampling induces a rigid displacement
of the family size distribution to the left and downward.   For strong but partial sampling the formula applies all the way down to $m=1$, whereas for the full population, the formula breaks down for the smallest $m$'s. For small argument, $U$ approaches a constant, and so for $m \ll s$, we get
\begin{equation}
n_m^R \approx \frac{\nu R_c}{ m}
\end{equation}
For large arguments, we recover the standard power-law.
This is evidenced in Fig. \ref{bigs}.

\begin{figure}
\center{\includegraphics[width=0.5\textwidth]{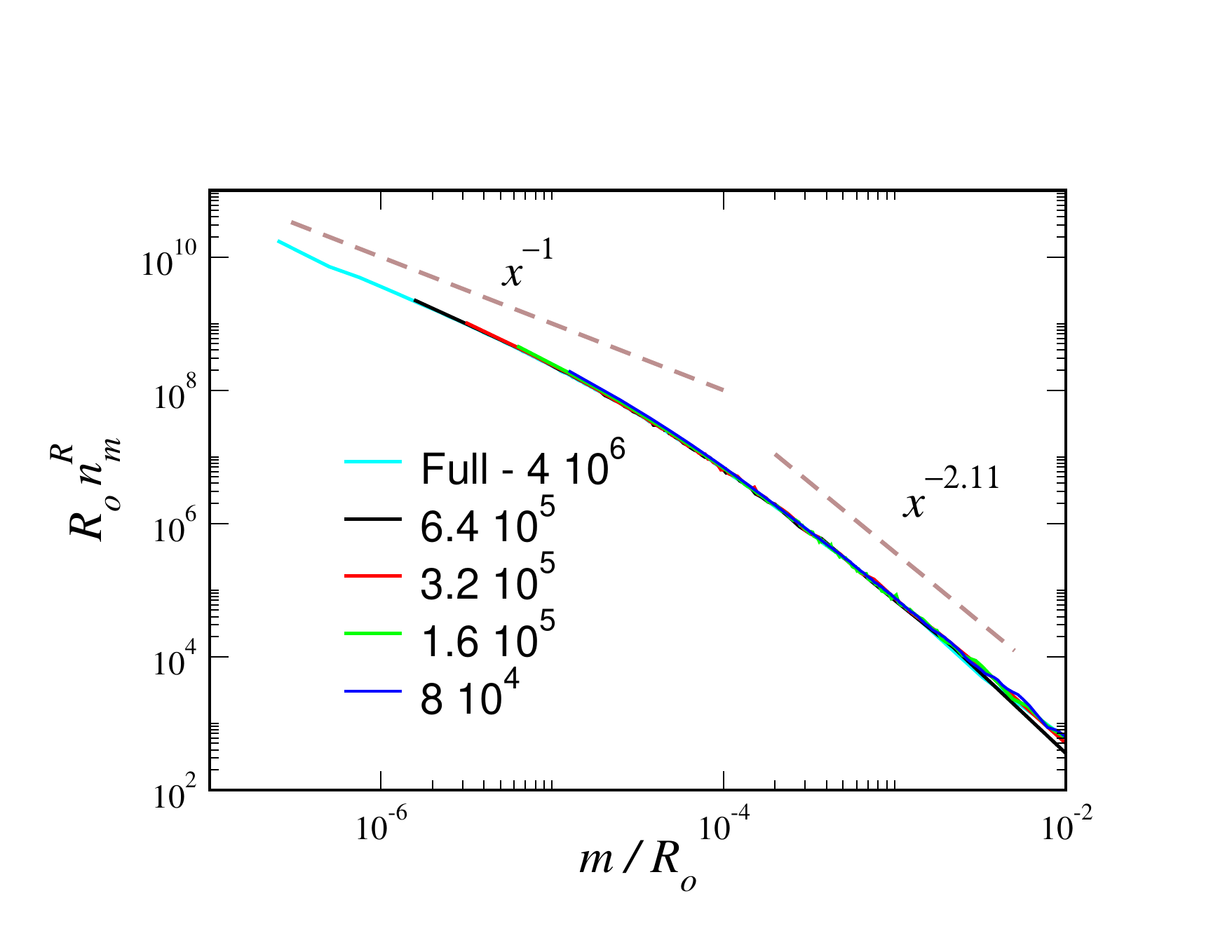}}
\caption{Data collapse for strong sampling: $R_o n^R_m$ as a function of $m/R_o$ for various sized samples, $R_o=6.4\cdot 10^5$, $3.2\cdot 10^5$, $1.6\cdot 10^5$ and $8\cdot 10^4$, along with the whole population $R_o=N_o=4 \cdot 10^6$.  Also shown are the small and large $s$ power law predictions.}
\label{bigs}
\end{figure}

For weak sampling, $R_o \ll \gamma N_o$, i.e., $s \ll 1$, the sampling strength decouples from the distribution for $m>1$ except to set the overall normalization:
\begin{equation}
n_m^R \approx \textrm{B}(2+\nu,m-1-\nu)\nu R_o s^\nu \qquad\qquad m>1
\label{dilute}
\end{equation}
This is demonstrated in Fig. \ref{dilutefig}.
The distribution rapidly approaches the expected power-law behavior from above as $m$ increases.
Thus, the shoulder has disappeared completely.  The families of family size 1, which we denote singletons, are exceptional for weak sampling, since the chance of sampling more than one member from a given family vanishes in the small $s$ limit, except for small (scaled) mutation rate $\nu$, where there are anomalously large numbers of large families.

\begin{figure}
\center{\includegraphics[width=0.5\textwidth]{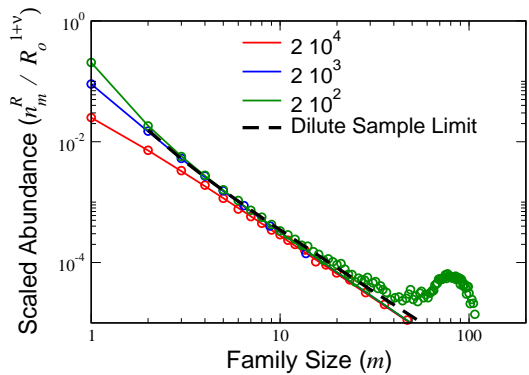}}
\caption{Data collapse for weak sampling: $n^R_m/R_o^{1+\nu}$ as a function of $m$ for various sized samples, $R_o=200$, $2000$ and $2\cdot 10^4$, along with the analytic prediction, Eq. (\ref{dilute}). }
\label{dilutefig}
\end{figure}

\item The average red family size, $\overline{m^R}$, is given by the equally monstrous formula
\begin{equation}
\overline{m^R}=\frac{s(1+\nu)}{\nu\Big[{}_2F_1\left(1,1;3+\nu;1\right)-(1-s)\,{}_2F_1\left(1,1;3+\nu;1-s\right)\Big]}
\end{equation}
This is shown in Fig. \ref{mbarfig}, where $\overline{m^R}$ is shown as a function of $R_o$, together with the results of numerical simulations.
For strong sampling, there of order $\ln s$  red families, and
\begin{equation}
\overline{m^R} \approx \frac{s}{\nu \ln a s}
\end{equation}
where $a$ is a $\nu$ dependent constant which approaches unity for small $\nu$, given by Eq. (\ref{aconst}).  In particular, in the full sample, $s=\sigma^2(1+\nu)/2\gamma$,
and the average family size is large,  of order $-1/(\gamma\nu \ln \gamma)$ 

\begin{figure}
\center{\includegraphics[width=0.5\textwidth]{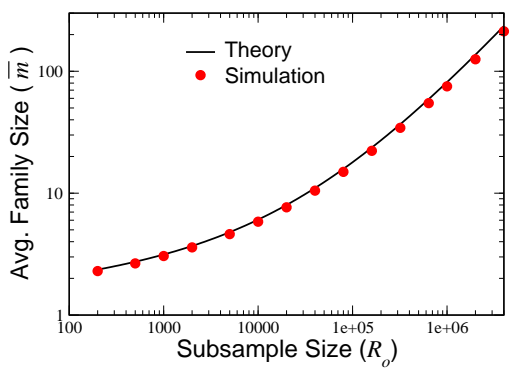}}
\caption{Average family size $\overline{m^R}$ as a function of the subsample size $R_o$ for $\gamma=0.005$, $\mu=5\cdot 10^{-4}$, and $N_o=4\cdot 10^6$.}
\label{mbarfig}
\end{figure}

For weak enough sampling, the average family size approaches unity, since all families are singletons.  For small $\nu$ however, this again occurs only for extremely small samples, and in practice,
\begin{equation}
\overline{m^R} \approx  - \frac{1-s}{\nu\ln s}
\end{equation}
so that for small $\nu$ the average red family size is large, and decreases logarithmically  as the sampling strength decreases.

\item  The distinction between weak and strong sampling is also reflected in the statistics of the genealogical tree of the subsample.  For strong sampling, this is a strong coalescence at first as we must upward in the tree.  For weak sampling, on the other hand, the coalescence is very small initially.  Eventually, both trees narrow exponentially.  For critical sampling, the narrowing of the tree is exponential for all times.

\end{enumerate}
This then is the general outline of our results.  In the next section, we turn to a detailed derivation of these findings.

\section{General analysis of the branching-mutation process}

We start with the family size distribution for the entire
population.  This is given by the solution to a Fokker-Planck
equation generalizing that derived by MZ for their model of
overlapping generations.  We start with the set of difference
equations for the evolution of the whole-population distribution:
\begin{subequations}
\begin{eqnarray}
\hspace*{-.25in}n_m^{t+1}\!\!\! &=&\!\!\! \sum_{\substack{\ell\\p\ge m}} n_\ell P(\ell \to p) {p \choose m} \mu^{p-m} (1-\mu)^m;\quad m>1\nonumber\\
&\ &\label{differenceeq}\\
\hspace*{-.25in}n_1^{t+1}\!\!\! &=&\!\!\! \sum_\ell n_\ell P(\ell \to p){p \choose
1}\mu^{p-1}(1-\mu)  + \mu N(t+1)\nonumber\\
&\ &
\end{eqnarray}
\end{subequations}
The first equation represents the contribution of a family of size
$\ell$ giving birth to a family of size $p$, $m-p$ are whom mutate,
leaving a family of size $m$.  The probability $P(\ell\to m)$ of
$\ell$ individuals to give birth to $m$ children is derived in term
of the fundamental distribution of the number of children of a
single individual, $P(m)$, whose mean we label $\lambda$.  All
mutations become new families of size one.

Before we present the differential equation, we can verify that
asymptotically for large $m$ (item 1 above), the stable distribution
falls like a power.  This can be accomplished by using the central
limit theorem and evaluating the sums via the Laplace method.  The
details are this calculation are presented in  Appendix \ref{app1}.
We find that indeed $n_m \sim m^{-\beta}$ where
\begin{equation}
\beta = \frac{\ln \lambda^2 (1-\mu)}{\ln \lambda(1-\mu)}
\label{betaeqn}
\end{equation}
This is different from the MZ result, and reflects the difference
between the overlapping generations in their model versus the
synchronized update of this model.  Nevertheless, writing the growth
factor $\lambda \equiv 1+\gamma$ and going to the slow growth, small
mutation  limit $\gamma \sim \mu \ll 1$, we get to leading order
\begin{equation}
\beta \approx \frac{2\gamma - \mu}{\gamma - \mu} = 2 + \nu
\end{equation}
This is the MZ result, indicating that for overlapping generations the growth rate is always small in some sense.  In any case, we conclude that when the growth and mutation
are sufficiently small so that the population changes slowly, the
details of the update procedure no longer matter.  

By using the generating function for the child distribution, we can
derive, as detailed in Appendix \ref{app2}, the Fokker-Planck
equation (item 2):
\begin{equation}
\frac{\partial n}{\partial t} = -(\gamma - \mu)
\frac{\partial}{\partial x} (xn) + \frac{\sigma^2}{2}
\frac{\partial^2}{\partial x^2} (xn) \label{MZeq}
\end{equation}
This equation is approximate.  In particular, the coefficients are
presented, in light of the discussion above, to leading order in
$\mu$ and  $\gamma$.  In addition, we have truncated the equation at
terms of second order.  There are additional second derivative
terms, which arise from the third ``spatial" derivative of $xn(x)$,
which we have as dropped, as do MZ. The first order derivative
represents the drift to larger population, with an effective growth
rate for the family of $\gamma-\mu$, since mutations reduce the
family size. The coefficient of the second derivative term,
$\sigma^2$, is the variance of the given children distribution of an
individual. It is eminently reasonable that the variance in the
number of children is what gives rise to diffusive behavior of the
family size.   For the case of the geometric distribution of
children, which characterizes the MZ model, with variance 2 in the
small $\gamma$ limit, the equation reduces to theirs.

Thus, up to the change of the diffusion constant, the stable
distribution in this  approximation is again the Kummer
function~\citep{AbSt}
\begin{equation}
n(x) \approx \frac{A}{x}
U\left(1+\nu,0,\frac{2(\gamma-\mu)}{\sigma^2}x\right)
\end{equation}
where $A$ is a normalization factor.  From here, we can justify the
neglect of the higher derivatives.  The argument of the Kummer
function is proportional to the small quantity $\gamma-\mu$, so that
higher derivatives bring down additional factors of this small
quantity and are thus indeed negligible.  This stems from the fact
that the drift term is small, and again hearkens back to the
necessity of assuming slow growth and small mutation probability.

The normalization factor can be obtained from the size of the entire
population, which we denote $N_o$, since
\begin{equation}
N_o = \sum_{m\ge 1} m n_m \approx \int_0^\infty x n(x)dx
\end{equation}
This integral can be performed analytically \citep[see][Eq. 7.612.2]{integrals},
 yielding (item 3)
\begin{equation}
n(x) =  \frac{\nu R_c}{x}\Gamma\left(2+\nu\right)\,
U\left(1+\nu,0,\frac{2\gamma}{\sigma^2(1+\nu)}x\right) \label{full}
\end{equation}
It should be noted that this normalization is different from that of
MZ. The fact that the distribution is proportional to $\mu$ flows
from the fact that for $\mu=0$ the distribution goes like $1/x^2$
and so  the integral diverges as $1/\mu$.

The form of the distribution means that $\sigma^4 n$ is a function
of the scaled family size $m/\sigma^2$, for all offspring
distributions.  This collapse is shown in Fig. \ref{scaledfs},
where the data for the Poisson and geometric offspring distributions
is plotted, together with the analytic prediction.  We see the
analytic prediction is excellent except for the singletons, the
families of size one.  In any case, these are not expected to be
given by the formula, since the governing equation for the
singletons is exceptional.

From the full population distribution, it is straightforward to
generate, at least formally, the family size distribution, $n^R_m$,
for the ``red" subsample of size $R_o$:
\begin{equation}
n^R_m = \sum_{p \ge m} n_p \frac{{p\choose m}{{N_o - p} \choose {R_o
- m}}}{{N_o \choose R_o}}
\label{hyper}
\end{equation}
which reflects the hypergeometric probability of choosing $m$ red
members of an original family of size $p$, when choosing $R_o$ from
$N_o$. When $R_o \ll N_o$, the hypergeometric distribution reduces
to a Poisson distribution
\begin{equation}
n^R_m\approx \sum_{p \ge m} n_p \frac{e^{-pR_o/N_o}}{m!}
\left(\frac{pR_o}{N_o}\right)^m
\label{sample}
\end{equation}
We can replace the sum by an integral, yielding \citep[see][Eq. 7.621.6]{integrals},  the result quoted
in the 4th item above, 
\begin{eqnarray}
\hspace*{-.25in}n^R_m &\approx& \frac{1}{m!}\left(\frac{R_o}{N_o}\right)^m\int_m^\infty dx\,x^{m-1} e^{-x R_o/N_o}\times\nonumber\\
&\ &\qquad AU\left(1+\nu, 0, \frac{2\gamma}{\sigma^2(1+\nu)} x\right)  \nonumber\\
&\approx& \nu R_c \,\textrm{B}\left(2+\nu,m\right)s^m\times\nonumber\\
&\ &\quad{}_2F_1\left(m,m+1;2+\nu + m; 1 - s\right)   .\label{subdist}
\end{eqnarray}
 In Fig.
\ref{fig2} above, we have included together with the simulational
results for the family size distribution for various size subsamples
the predictions of our formula Eq. (\ref{subdist}). We see that for
all but the largest families in the smallest subsample, there is
excellent agreement, even for a subsample which is $1/20000$ of the
whole.

The first thing we can verify with our analytic formula is that the
large $m$ behavior of the distribution is unchanged, except for
normalization. For large $m$, the sum is dominated by $p$'s of order
$m N_o/R_o$.  Expanding the summand around this maximum yields a
Gaussian, and we see that power-law is preserved.

We now investigate our result in the two limits of strong and weak sampling, defined by
$s\gg 1$ and $s\ll 1$, respectively.  In the limit of strong sampling, we return to  our fundamental expression for $n_m^R$, Eq. (\ref{sample}).  As we shall see momentarily, the typical scale of family size over which $n_m$ varies is large, of order $s$.  Thus, the Poisson sum is essentially a Gaussian centered at $\ell^*=mN_o/R_o$, of width $\sqrt{\ell^*}$. Since $n_m$ itself varies over the scale $1/\gamma$, which is much larger, to leading order the
sum over $\ell$  reduces to a $\delta$-function, and we have
\begin{equation}
n_m^R \approx \nu R_c\frac{\Gamma(2+\nu)}{m}U(1+\nu,0,m/s)
\end{equation}
Alternatively, we can obtain this expression directly from our general formula for $n_m^R$ using the integral representation of ${}_2F_1$ and taking the large $s$ limit.
The first derivation shows that the result is valid  even for $R_o \sim N_o$, beyond the range of the Poisson sampling approximation, since the argument generalizes to the original hypergeometric sampling formula, Eq. (\ref{hyper}).  This is evident from the fact that we recover the full population distribution if we simply set $R_o = N_o$.  However, the partial sample distribution formula is reliable for all $m$, whereas for the full population, the formula does not apply to $m$'s of order unity.  As we noted above, this form for $n_m^R$ implies that as we vary $R_o$, the whole
distribution moves rigidly leftward and downward.  In particular, for $m \ll s$, $U$ approaches a constant value, $1/\Gamma(2+\nu)$, and
\begin{equation}
n_m^R \approx \frac{\nu R_c}{m}
\label{strongsmallm}
\end{equation}
Clearly for $m \gg s$, we recover the MZ power-law, as expected.  We refer the reader back to Fig. \ref{bigs} for a graphical presentation of the strong sampling regime.  

Since the shoulder of the distribution extends to $m$'s of order $s$, clearly the shoulder vanishes by the time $s$ reaches 1.  We call this "critical" sampling, and in this case the
distribution is given by
\begin{equation}
n_m^R = \nu R_c\, \textrm{B}\left(2+\nu,m\right)
\end{equation}
In particular, for small $\nu$, this reduces to the simple form
\begin{equation}
n_m^R = \frac{\nu R_c}{m(m+1)}
\end{equation}
and show that in this case the distribution still approaches the large $m$ power law from below, but the power-law regime sets in already for $m \gtrsim 5$ or so.

We now turn to the weak sampling regime, $s \ll 1$. Here, using the transformation formula~\citep{AbSt}
\begin{eqnarray}
\hspace*{-.25in}{}_2F_1(a,b;c;z)\!\!\! &=& \frac{\Gamma(c)\Gamma(c-a-b)}{\Gamma(c-a)\Gamma(c-b)}{}_2F_1(a,b;a+b-c+1;1-z) \nonumber\\
&\ & {} + (1-z)^{c-a-b}\frac{\Gamma(c)\Gamma(a+b-c)}{\Gamma(a)\Gamma(b)}\times\nonumber\\
&\ &\qquad{}_2F_1(c-a,c-b;c-a-b+1;1-z) \nonumber\\
&\ &
\end{eqnarray}
we note that for $m>1$, the second term dominates for small $s$.  Thus, we get to leading order
\begin{equation}
n_m^R \approx \textrm{B}(2+\nu,m-1-\nu)\nu R_o s^\nu \qquad\qquad m>1
\end{equation}
so that the $m>1$ distribution is independent of $R_o$, up to normalization.  We see that as $s \to 0$, $n_m^R/R_o$ vanishes for $m>1$, since extremely dilute sampling will never encounter two individuals from the same family.  Again, the small $\nu$ limit is particularly simple
\begin{equation}
n_m^R \approx \frac{\nu R_o}{m(m-1)}  \qquad\qquad m>1
\end{equation}
We see clearly that the case $m=1$ is exceptional, and requires a separate treatment.  We also see that the distribution now approaches the asymptotic power-law from above.
The weak sampling regime is illustrated in Fig. \ref{dilutefig} above.

We have seen that the case of the singleton families, $m=1$, requires special attention,
particularly for weak sampling.
Using the transformation formula above, we can verify that in the weak sampling limit, $s \to 0$,  $n_1^R$ approaches $R_o$, consistent with the vanishing of the larger size families in this limit.   However, this is
misleading for small $\mu$.  In this limit, the hypergeometric
function can be expressed in terms of elementary functions, and
\begin{equation}
n_1^R \approx \frac{-R_o \nu}{(1-s)^2}(1-s+\ln s) +
{\cal{O}}(\nu^2) \label{rs}
\end{equation}
Thus, for dilute sampling, $n_1^R/R_o$ is seen to be approximately
the small quantity $-\nu \ln(s)$,  The logarithmic behavior
is again a reflection of the slow $1/x^2$ decay of the whole
population family distribution in the small $\mu$ limit. Only for
ridiculously dilute samples, with $R_o/N_o$ of order $\gamma
e^{-\gamma/\mu}$, does the whole subsample reduce to singletons.
This is what we see in the graphs of the simulations in Fig.
\ref{fig2}.where even for the smallest subsample shown, with
$s=0.0222$, there are only 73 singletons in a population of 200.
In the strong sampling regime, as can be seen from Eq. (\ref{strongsmallm}), 
we get $n_1^R \approx \nu R_c$, so it is independent of
$R_o$. This convergence is also apparent in the data.  Overall,
 the fraction of singletons decreases with the strength of the
sampling.

The last quantity of interest is the total number of red families,
which we denote by $F^R$.  In principle we can calculate it as the
sum over $m$ red family size distribution $n_m^R$, but it is easier
to go back to the definition of $n_m^R$ in terms of $n_p$ and sum
over $m$ first, leaving the sum over $p$ for last.  The sum over
$m$, if it started at $m=0$ would just give unity, but because it
starts at $m=1$, yields, in the Poisson approximation $R_o \ll N_o$:
\begin{equation}
F^R\approx \sum_{p \ge 1} n_p \left(1 - e^{-pR_o/N_o}\right)
\end{equation}
Thus, all families in the full population give a family of
{\emph{some}} size in the red population, unless they are not picked
at all, which occurs with probability $exp(-p R_o/N_o)$, where $p$
is the size of the originating family.  Plugging in our expression
for $n_p$, and converting the sum to an integral yields
\begin{subequations}
\begin{eqnarray}
F^R&\approx& \int_0^\infty\left(1 - e^{-xR_o/N_o}\right)\times\nonumber\\
&\ &\qquad\qquad \frac{A}{x} U\left(1+\nu,0,\frac{2\gamma}{\sigma^2(1+\nu)}x\right) dx\nonumber\\
&\ &\label{rf}\\
&=& \frac{\nu R_c}{\left(2+\nu\right)}
\Big[{}_2F_1\left(1,1;3+\nu;1\right) - \nonumber\\
&\ &\qquad \quad(1-s)\,{}_2F_1\left(1,1;3+\nu;1-s\right)\Big]\nonumber\\
&\ &
\end{eqnarray}
\end{subequations}
The details of this calculation are presented in Appendix
\ref{families}.  Again, formally, for any finite $\nu$, we get  that
$F^R$ approaches $R_o$ as $s \to 0$, as all families are singletons.
Nevertheless, 
 for small, but not absurdly small,  $\nu$, the number of families reads
\begin{equation}
F^R\approx -\nu R_c \frac{s\ln s}{1-s} \ . \label{rflownu}
\end{equation}
Thus, the average family size, $\overline{m}\equiv R_o/F^R$ is given for small $\nu$ by
\begin{equation}
\overline{m} = - \frac{1-s}{\nu\ln s}\label{frnu}\end{equation}
which is large but decreases logarithmically at small sampling, cutting off at one for extremely small samples.  
density $s$. It is interesting that whereas both the number of red
families and the number of red singletons behave anomalously for
small samples and small $\mu$, the fraction of red families that are
singletons is smooth, approaching unity as it should. 

At critical sampling, $F^R = \nu R_c/(1+\nu)$, and $\overline{m} = (1+\nu)/\nu$, which is large for small $\nu$. 
For large $s$, and general $\nu$, the number of families is dominated by the contribution from the small families, which behaves as $1/m$ and so is logarithmically large:
\begin{equation}
F^R \approx \nu R_c\ln as
\end{equation}
where the ${\cal{O}}(1)$ constant $a$ is given by
\begin{equation}
\ln a= \psi(1) - \psi(2+\nu) + \frac{1}{1+\nu}
\label{aconst}
\end{equation}
and $\psi$ is the digamma function, so that $a$ approaches 1 for small $\nu$.
Thus the average family size diverges for large $s$,
\begin{equation}
\overline{m} \approx \frac{s}{\nu \ln as}
\end{equation}
in agreement with what we found for small $\nu$. 

\section{Red Statistics Through Time}
\subsection{Red Population}
We now present an alternative derivation of these results (at least
in the small $\nu$ limit), based on tracing the time development of
the red genealogy. We have labeled an individual ``red" if he is in
the selected sample of the final population.  We also label as red
any ancestor of such an individual.  We first address the question
of the time development of the red population.  The basic tool for
the analysis, as with the original Galton-Watson work, is the
generating function for the distribution of children, which we
denote by
\begin{equation}
F(x)\equiv \sum_{n=0}^\infty P(n) x^n
\end{equation}
The key is the Galton-Watson observation that the generating
function for the probability of descendants in the second
generation, $F_2(x)$ is the second iterate of $F$; i.e., $F(F(x))$.
We can see this noting that
\begin{equation}
P_2(n) = \sum_{i=0}^\infty P(i)P^i(n)
\end{equation}
so that
\begin{eqnarray}
F_2(x)&=&\sum_{i,n} P(i) P(i\to n) x^n = \sum_i P(i)[F(x)]^i\nonumber\\
&=& F(F(x))
\end{eqnarray}
.  Generalizing, the generating function for the probability of
descendants in the $n'th$ generation, $F_n(x)=F(F_{n-1}(x))$.

We now need to include the effects of sampling.  We ask what the
probability $Q_n$ that a person $n$ generations before the end has
no red descendent in the final sample.  This is just the sum of the
probabilities that it had $k$ descendants, and that none of these
were picked to be in the sample:
\begin{equation}
Q_n = \sum_{k=0}^\infty P_n(k) \left(1 - \frac{R_o}{N_o} \right)^k =
F_n \left(1 - \frac{R_o}{N_o}\right)
\end{equation}
where the final population is $N_o$ and the sample size is $R_o$.
The red population $n$ generations before the end is then
\begin{equation}
R_n = N_o \lambda^{-n} (1 - Q_n) = N_o \lambda^{-n} \left(1 - F_n
\left(1 - \frac{R_o}{N_o}\right)\right)
\end{equation}
This is the exact answer.  The function $F_n$ is what appears in the
Galton-Watson theory, and $F_n(x)$ approaches the
Galton-Watson-Steffensen fixed point (which exists for all
$\lambda>1$) for all $x<1$. This implies there is an interesting
change of behavior of $R_n$ depending on whether $1 - R_o/N_o$ is
greater, smaller or equal to  the fixed point value.  At the fixed
point, the percentage of reds in the general population is constant
as a function of time.  For a smaller sample, the ratio increases as
we move back in time, starting from the small initial value.  For a
larger sample, the situation is reversed, and the ratio decreases to
the asymptotic Galton-Watson survival probability as we go back in
time.

In order to get a more useful expression, we will specialize to our
limit of $\lambda$ close to 1, i.e.,   $\gamma \ll 1$.  We first
calculate the Galton-Watson (GW) fixed point in this limit. Since
for zero growth, the GW fixed point is unity, it is close to this
for $\gamma$ small. Writing $Q_\infty=1-\delta$,  the
fixed-point equation $Q_\infty = F(Q_\infty)$ reads
\begin{equation}
1-\delta = F(1-\delta) \approx 1 - \delta \lambda +
\frac{\delta^2}{2} \left[\sigma^2 - \lambda(1-\lambda)\right]
\end{equation}
where as before $\sigma$ is the variance of the children
distribution, so that
\begin{equation}
\delta \approx \frac{2\gamma}{\sigma^2}
\end{equation}
We thus see the origin of the ``critical" value of sampling we encountered in the previous section, namely the change in behavior depending on whether $R_o/N_o$ is less than or greater than $\delta \approx 2\gamma/\sigma^2$; i.e., on whether
$R_o$ is less than or greater than $2\gamma N_o/\sigma^2$, which is $R_c$ to leading order in $\nu$.  For the remainder of this section, we will in fact rewrite
$2\gamma N_o/\sigma^2 = R_c$, as we are working only to leading order in $\nu$.

As long as  $x$ is close to one, $F(x)$ will similarly be
close to one.  Thus, for $R_o \ll N_o$, $1-R_o/N_o$ meets this
criterion and we can approximate the change in $\delta_n =
R_n/N_n=1-Q_n$, the fraction of reds in the population.
\begin{eqnarray}
\delta_{n+1} &=& 1 - F(1-\delta_n) \approx \lambda \delta_n -
\frac{\delta_n^2}{2}\left(\sigma^2 - \lambda(1-\lambda)\right)\nonumber\\
&\approx& (1+\gamma)\delta_n - \frac{\delta_n^2\sigma^2}{2}
\end{eqnarray}
so that
\begin{equation}
\frac{d\delta}{dn} = \gamma \delta - \frac{\sigma^2\delta^2}{2}
\end{equation}
with the solution
\begin{equation}
\delta_n = \frac{2 \gamma R_o}{R_o \sigma^2 + (2 \gamma N_o -
R_o\sigma^2)e^{-\gamma n}}
\end{equation}
 so that
\begin{equation}
R(n) = \frac{R_c}{R_o  e^{\gamma n} + (R_c - R_o)} R_o
\label{red}
\end{equation}
The change in behavior as $R_o$ crosses $R_c$ is apparent. In particular, at critical sampling, $R(n)$ is a pure exponential in time.  We also see that $R_n$ depends on the
underlying distribution of children only through its average, i.e.
$\gamma$, and its variance.

\begin{figure}
\center{\includegraphics[width=0.5\textwidth]{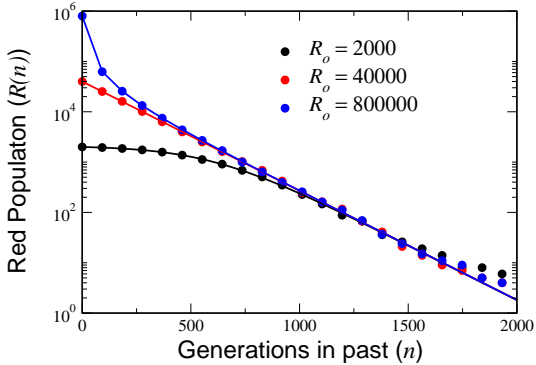}}
\caption{The number of red individuals as a function of the number of generations in the past, $R(n)$, for a \emph{single} run, with $N_o=4\cdot 10^6$, $\gamma=5\cdot 10^{-3}$, $\mu=5\cdot 10^{-4}$.  The sample sizes are $R_o = 2000$, $4\cdot 10^4$, and $8\cdot 10^5$, which are subcritical, critical and supercritical respectively.  Also shown as solid lines are the predictions of Eq. (\ref{red}).}
\label{red_t_fig}
\end{figure}

In Fig. \ref{red_t_fig}, we show data for the $R(n)$ from a single simulation, for three different sampling strengths, together with our analytic formula, Eq. (\ref{red}).  The data exhibit the striking change in behavior from a fast initial decrease in red individuals for strong sampling, a pure exponential for critical sampling and a slow initial decrease for weak sampling.  All three curves merge in the past, at the coalescence time for the entire population.  

An alternate, equivalent way to arrive at our result is to consider
the coalescence of branches on the red tree.  If we look at the
$R_n$ reds in the $n$th previous generation, these are generated by
slightly less than $R_n$ parents, due to coalescence.  The chances
of coalescence, per potential parent,  is the chance of having two
surviving children, $\sum P(n) \frac{n(n-1)}{2}\delta^2 \approx
\sigma^2\delta^2/2$.  Thus, the decrease in the number of reds is
$\sigma^2 R^2/2N$ so that the equation for $R$ is
\begin{equation}
\frac{dR}{dn} = - \frac{\sigma^2 R^2}{2N}
\end{equation}
which is equivalent to our above equation for $\delta$.

\subsection{Red Families}
We now turn to examine the time development of the number of red
families, $F^R(n)$, $n$ steps in the past . As in the absence of
mutation every family survives, since it is red, the only change in
families from $n$ steps in the past to $n-1$ steps in the past is
the new families due to mutation.  As mutations of singletons do not
create new families, we have
\begin{equation}
F^R(n-1)=F^R(n) + \mu (R(n) - n_1^R(n))\ .
\end{equation}
In the small $\mu$ limit, the number of singletons is small,
proportional to $\mu$.  Thus to leading order in $\mu$, we have
\begin{equation}
F^R(n-1)=F^R(n) + \mu R(n)
\end{equation}
or, in its differential equation form,
\begin{equation}
\frac{dF^R}{dn} = -\mu R
\end{equation}
Using our solution, Eq. (\ref{red}), for $R(n)$, we obtain
\begin{equation}
F^R(n) = \frac{\nu R_c R_o}{R_c - R_o}\ln\left(1 +
\left( \frac{R_c}{R_o} - 1\right)e^{-\gamma
n}\right)
\label{frn}
\end{equation}
where we have demanded that $F^R \to 0$ as $n\to \infty$.  In
particular, at the sampling time,
\begin{equation}
F^R(0) = \frac{\nu R_c R_o}{R_c - R_o}\ln
\frac{R_c}{R_o}
\end{equation}
which of course agrees with the small $\mu$ limit result Eq.
(\ref{frnu}) we derived using the Fokker-Planck approach.

\begin{figure}
\center{\includegraphics[width=0.5\textwidth]{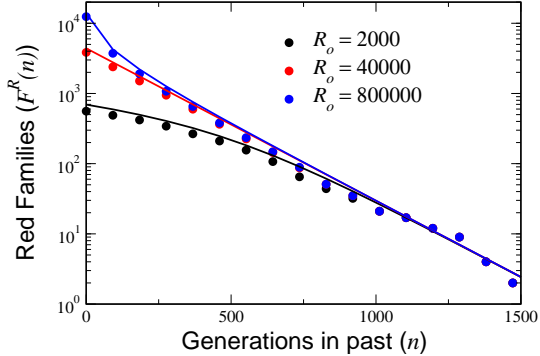}}
\caption{The number of red families as a function of the number of generations in the past, $F^R(n)$, for a \emph{single} run, with $N_o=4\cdot 10^6$, $\gamma=5\cdot 10^{-3}$, $\mu=5\cdot 10^{-4}$.  The sample sizes are $R_o = 2000$, $4\cdot 10^4$, and $8\cdot 10^5$, which are subcritical, critical and supercritical respectively.  Also shown as solid lines are the small $\nu$ approximations from Eq. (\ref{frn}).}
\label{redfam_t_fig}
\end{figure}

In Figure \ref{redfam_t_fig}, we show data for the number of red families going backward in time, again for subcritical, critical and supercritical sampling.  The
small $\nu$ result, Eq. (\ref{frn}), is also shown.  The small deviation is due to higher order corrections in $\nu$. 

\subsection{Red Singletons}
\begin{figure}
\center{\includegraphics[width=0.5\textwidth]{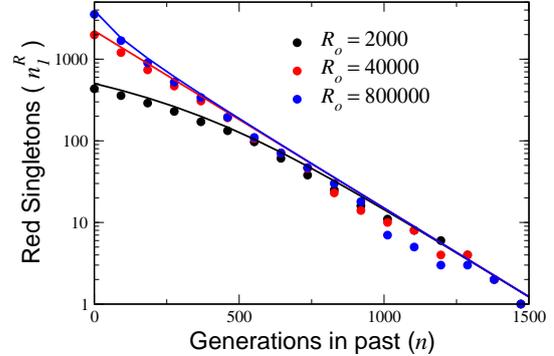}}
\caption{The number of red singletons as a function of the number of generations in the past, $n_1^R(n)$, for a \emph{single} run, with $N_o=4\cdot 10^6$, $\gamma=5\cdot 10^{-3}$, $\mu=5\cdot 10^{-4}$.  The sample sizes are $R_o = 2000$, $4\cdot 10^4$, and $8\cdot 10^5$, which are subcritical, critical and supercritical respectively.  Also shown as solid lines are the small $\nu$ approximations from Eq. (\ref{sing}).}
\label{redsing_t_fig}
\end{figure}

We now examine  the time development of the number of red
singletons.   The number of singletons decreases due to the fact
that some singletons give birth to multiple children, who are no
longer singletons.  It increases due to the fact that mutations of
non-singletons give rise to new singletons.  The latter factor is
very simple:  $\mu(R - n_1^R)$, just as with red families.  As
discussed above, the  decreases in number of reds due to coalescence
is $\sigma^2 R(n)^2/2N(n)$.  Of these a fraction $n_1^R/R$ are
singletons, so the loss of singletons due to coalescence of
singletons  is $\sigma^2 R n_1^R/2N$.  Thus the differential
equation for $n_1^R$ is
\begin{equation}
\frac{dn_1^R}{dn} = \frac{\sigma^2 Rn_1^R}{2N_o e^{-\gamma n}} - \mu
(R - n_1^R)
\end{equation}
Again we can drop the $\mu n_1^R$ term as being higher order in
$\mu$.  Then the solution that vanishes as $n\to \infty$ is
\begin{eqnarray}
n_1^R(t) &=& \frac{\nu R_c R_o }{(R_c -
R_o)^2}\Bigg[ -(R_c-R_o) + \nonumber\\
&\ &\Big(R_o e^{\gamma n} + \left(R_c -R_o\right) \Big)\times\nonumber\\
&\ &\qquad\ln\left(1 + \left(\frac{R_c}{R_o}-1\right)e^{-\gamma n} \right)\Bigg] \nonumber\\
\label{sing}
\end{eqnarray}
In particular, at the sampling time the number of red singletons is
given by
\begin{equation}
n_1^R(0) = \frac{\nu R_c R_o}{(R_c - R_o)^2} \left[
R_c \ln \frac{R_c}{R_o} - (R_c -
R_o)\right]
\end{equation}
This again can be seen to reproduce our Fokker-Planck result, Eq.
(\ref{rs}).
  It is interesting to note that at critical sampling  $n_1^R$ approaches $\nu R_c/2$, whereas $F^R$ approaches $\nu R_c$, so that half the families are singletons in this case.  The simulation data for $n_1^R(n)$ is presented in Fig. \ref{redsing_t_fig}, together with the small $\nu$ prediction, Eq. (\ref{sing}).  The picture is qualitatively similar to that of the red families.
 
\section{The U.S. Census Data}

We now attempt to apply the theory to the surname distribution taken from the 2000
U.S. Census\footnote{Available at http://www.census.gov/genealogy/www/freqnames2k.html. This data only extends to surnames with more than 100 representatives.  Data for rarer surnames is available in binned form from http://www.census.gov/genealogy/www/surnames.pdf.  The binned data was then debinned using a smoothing procedure}.  We must make clear at the outset that degree of overlap between the assumptions underlying our theoretical treatment and the real dynamics of the surname distribution may rightfully be questioned.  Nevertheless, we will proceed and see how far we can go. 

\begin{figure}
\center{\includegraphics[width=0.5\textwidth]{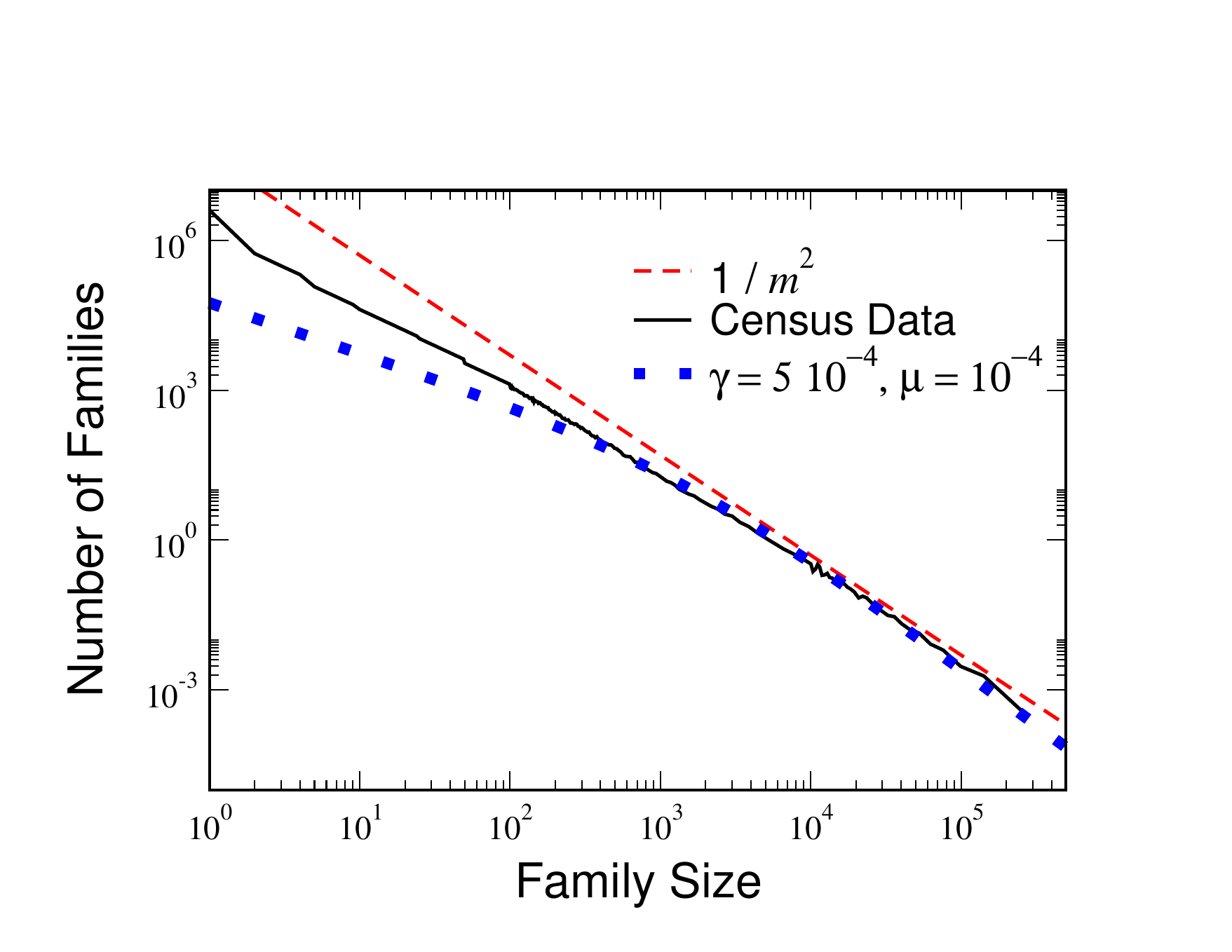}}
\caption{The surname distribution taken from the 2000 U.S. census.  Also show is a curve outlining a $1/m^2$ power-law decay, and the theoretical curve, Eq. (\ref{fulln}) for $N_o=2.8\cdot 10^8$, $\gamma=5\cdot 10^{-4}$, and $\mu=10^{-4}$.}
\label{usadist}
\end{figure}

The surname data is presented in Fig. \ref{usadist}.  We see that its overall shape is similar to that of the theory.  In particular, the graph exhibits a $1/m^2$
falloff for large family sizes, in accord with our expectation.  However, upon closer examination, the data presents us with a severe problem.  The asymptotic power law only sets in for family sizes above $10^4$ or so.  According to the theory, the onset of the power law should occur roughly at $m\sim 10/(2\gamma)$.  Thus, the data would point to a value of $\gamma$ of roughly $5\cdot 10^{-4}$.  However, this is off from the true growth rate of the population by a factor of 1000! 

\begin{figure}
\center{\includegraphics[width=0.5\textwidth]{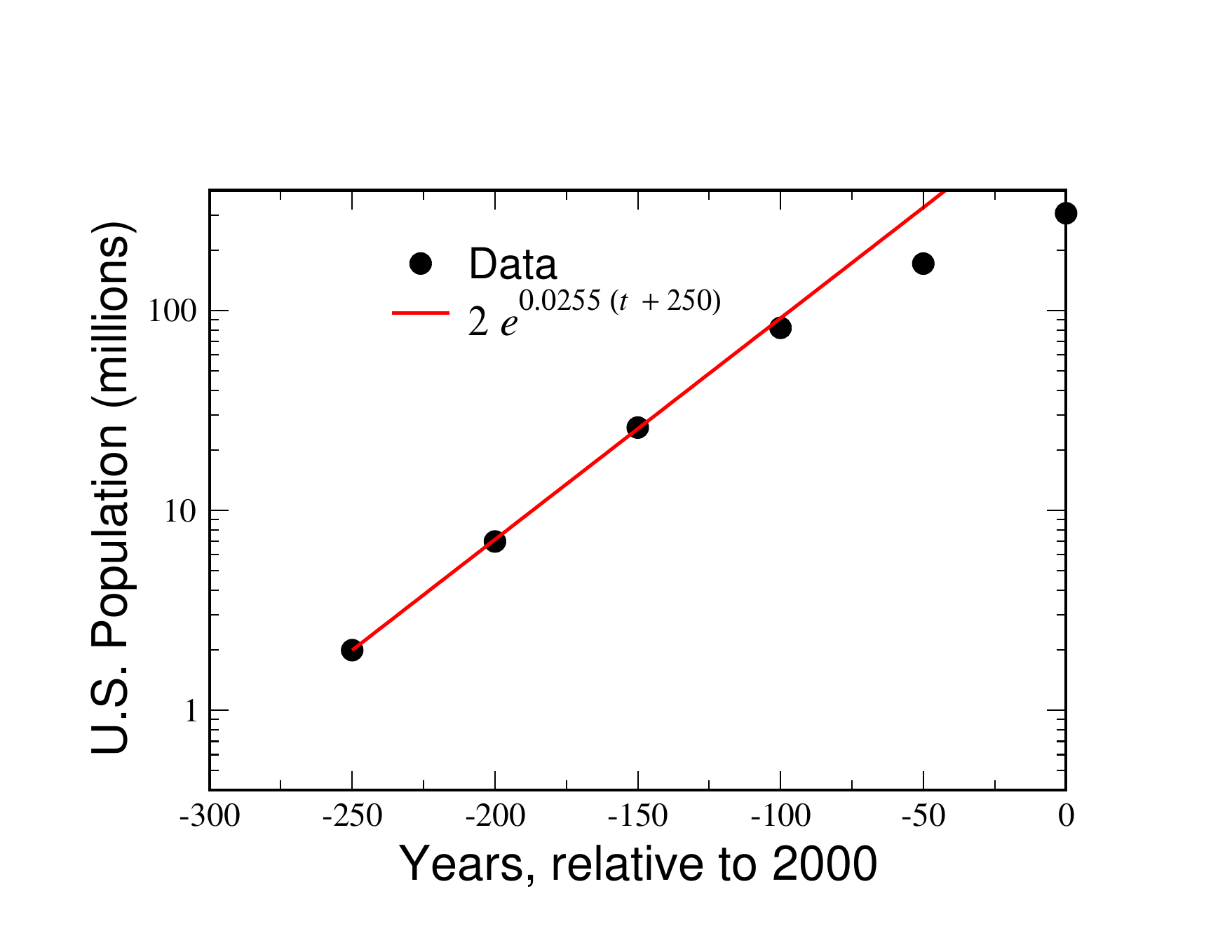}}
\caption{The population of the U.S. as a function of time, together with an exponential fit to the period prior to 1920}
\label{uspoptfig}
\end{figure}

The growth rate of the U.S. population has not been constant in time.  However, it was quite constant up until the severe immigration restrictions were applied in the wake of World War I, as can be seen from Fig. \ref{uspoptfig}, with a value of $0.0255\!\textit{/year}$. Assuming a generational time of 20 years, this works out to 
give a value of $\gamma=0.51$.  Even taking the post-WWI growth rate, the growth rate is only reduced to $\gamma=0.3$, so we still have a factor of 500 to account for. Even if we compare the data to the theoretical curve with $\gamma=5\cdot10^{-4}$, so as to reproduce the break from the power-law at $m=10^4$, the agreement only extends down to family sizes of $m=500$.

One might think that the problem is that the U.S.  is not a demographically closed system.  The large impact of immigration in driving the population growth before WWI is clear evidence of this.  One might then want to claim that the U.S. population should be considered as a   (biased) sample of the European (or perhaps, world) population.  The growth rate of the European and world population before 1920 is roughly a factor of 5 smaller than the U.S. growth rate.  However,
this improvement is more than counterbalanced by the sampling effect, which is to move the distribution leftward by the sampling factor, thereby moving the onset of the asymptotic $1/m^2$ scaling to even smaller $m$.  Thus, the solution cannot lie in this direction.

In fact, no clear answer presents itself at this point, and we must leave this puzzle as a challenge for the future.  One possibility might be that the "mutation" of family names, at least for the U.S., is likely very different from that assumed by the model.  There was a great tendency in the period of the great immigration for family name changes not to create new surnames, as we assume, but in fact the opposite.  The immigration officials were (in)famous for mapping the wide spectrum of surnames they encountered onto recognizable "American" names.  Further work will be needed to see what effect this might have had on the surname distribution.

In summary, then, we have calculated the family size distribution for a growing population.  We have shown that the distribution is universal for slow growth rate.  In addition, we have calculated the distribution for arbitrarily sized subsamples of the  population. We found a distinction between
strong sampling, where the shoulder at small family sizes, while
truncated still exists, and weak sampling, where the small family
distribution lies above the power-law.  The critical sampling
dividing these two regimes is $R_c=2\gamma N_o/\sigma^2(1+\nu)$.    In the strong sampling regime, the distribution is rigidly translated (in log space) through a rescaling of $m$.  For weak sampling, the distribution is independent of sampling, up to overall normalization.  We have also seen that the distinction between weak and strong sampling holds for the properties of the genealogical tree constructed from the sampled individuals.

 This work was supported by the EU 6th framework CO3 pathfinder.
 Y.M. acknowledge the financial support of the Israeli  Center for
Complexity Science.

\appendix\section{Derivation of the Power-Law\label{app1}}
For large $m$, the major contribution of the sum over $p$ is from
the vicinity of $p_*\equiv m/(1-\mu)$ and for the sum over $\ell$
from the vicinity of $\ell_* \equiv p_*/\lambda =
m_*/((1-\mu)\lambda)$. Thus, we can, invoking the central limit
theorem, approximate the distribution $P(\ell\to p)$ by
\begin{equation}
P(\ell\to p) \approx \frac{1}{\sqrt{2\pi\sigma^2 \ell}}
e^{-\frac{(p-\lambda \ell)^2}{2\sigma^2}}
\end{equation}
Similarly, we can approximate the binomial distribution for the
number of mutations by
\begin{equation}
{p \choose m} (1-\mu)^m \mu^{p-m} \approx
\frac{1}{\sqrt{2\pi\mu(1-\mu) p}}
e^{-\frac{(m-p(1-\mu))^2}{2\mu(1-\mu)p}}
\end{equation}
Replacing the sums over $\ell$ and $p$ by integrals and expanding
the exponent to second order yields
\begin{eqnarray}
n_m^{t+1} &\approx& \iint d\ell\, dp\  n_\ell^t \frac{1}{2\pi\sigma\sqrt{\ell p \mu(1-\mu)}}\times\nonumber\\
&\ & \exp\left(-\frac{(p-\lambda \ell)^2}{2\sigma^2\ell}-\frac{(m-p(1-\mu))^2}{2\mu(1-\mu)p}\right)\nonumber\\
&\approx&\frac{\sqrt{\lambda(1-\mu)}}{2\pi\sigma m \sqrt{ \mu}}
n_{\ell_*}^t \iint d\ell\, dp\   e^{{\cal{F}}(\ell,p)}
\end{eqnarray}
where
\begin{eqnarray}
{\cal{F}}(\ell,p)&=&-\frac{\lambda^3(1-\mu)(\ell -
m_*)^2}{2\sigma^2 m}\nonumber\\
&&{}-\frac{(1-\mu)(\lambda\mu+\sigma^2(1-\mu))(p-p_*)^2}{2\mu m}\nonumber\\
&\ &{}-
\frac{\lambda^2(1-\mu)(\ell-\ell_*)(p-p_*)}{\sigma^2 m}
\end{eqnarray}
Substituting $n_m^t = Am^{-\beta}\lambda^t$ and doing the Gaussian
integrals gives
\begin{equation}
\lambda m^{-\beta} =\frac{1}{\lambda(1-\mu)}
\left(\frac{m}{\lambda(1-\mu)}\right)^{-\beta}
\end{equation}
Equivalently, we could simply replace the Gaussians by the
$\delta$-functions $\delta(p-m\ell)\delta(m - (1-\mu) p)$ and
integrate over $\ell$ and $p$ successively. Taking the logarithm
gives us our equation for $\beta$, Eq. (\ref{betaeqn}).

\section{The Fokker-Planck Equation\label{app2}}
We start with Eq. (\ref{differenceeq}), and write $P(\ell\to p)$ in
terms of the generating function, $F$:
\begin{equation}
P(\ell\to p) = \oint \frac{dz}{z^{p+1}} F^\ell(z)
\end{equation}
giving
\begin{equation}
n_m^{t+1} = \sum_{\substack{\ell\ge 0\\p\ge m}} \oint
\frac{dz}{z^{p+1}} F^\ell(z) n_\ell^t {p \choose
m}\mu^{p-m}(1-\mu)^m
\end{equation}
We next expand $n_\ell^t$ in a Taylor series around $n_m^t$:
\begin{equation}
n_\ell^t = n_m^t + \frac{\partial n}{\partial m} (\ell-m) +
\frac{1}{2}\frac{\partial^2 n}{\partial m^2}(\ell - m)^2 + \ldots
\end{equation}
We can now do the geometric sums over $\ell$ and the sum over $p$
using
\begin{equation}
\sum_{p \ge m} {p \choose m} x^p = \frac{x^m}{(1-x)^{m+1}}
\end{equation}
to get
\begin{eqnarray}
n_m^{t+1} &=& \oint \frac{dz}{2\pi i} \frac{(1-\mu)^m}{(z-\mu)^{m+1}} \Bigg[ \frac{n - n' m + n'' m^2/2}{1 - F(z)} \nonumber\\
&\ &{}+ \frac{(n' - mn'')F(z)}{(1-F(z))^2} + \frac{(n''/2)F(z)(1+F(z))}{(1-F(z))^3}\Bigg] \nonumber\\
&=& \frac{n-n' m + n''m^2/2}{\lambda(1-\mu)}\nonumber\\
&\ &{}+ \frac{(n' - mn'')\left[(\lambda^2-f_2)(1-\mu)+(m+1)\lambda\right]}{\lambda^3(1-\mu)^2}\nonumber \\
&\ &{}+ \frac{n''/2}{\lambda^5(1-\mu)^3}\Bigg[-6\lambda(1-\mu)^2f_3 \nonumber\\
&\ &{}\qquad+ 12(1-\mu)^2f_2^2 - 6\lambda(1-\mu)^2\lambda f_2 \nonumber\\
&\ &\qquad{}+ 6\lambda(1-\mu)(m+1)f_2 + \lambda^4(1-\mu)^2\nonumber\\
&\ &\qquad{} - 3\lambda^3
(1-\mu)(m+1) + (m+2)(m+1)\lambda^2\Bigg]\nonumber\\
&\ &
\end{eqnarray}
where we have set the contour between the pole at $z=\mu$ and that
at $z=1$ and picked up the residue at the outer pole at $z=1$, and
the expansion of $F(z)$ near $z=1$ is
\begin{equation}
F(z) \approx 1 + \lambda (z-1) + f_2(z-1)^2 + f_3(z-1)^3 + \ldots
\end{equation}
It is possible to translate this difference equation to a
differential equation only in the limit $\gamma \sim \mu \ll 1$.
Then, expanding the coefficients of the various derivatives of $n$
to leading order, things simplify to
\begin{equation}
\dot{n}(m) = -(\gamma - \mu) n +\left[-(\gamma-\mu)m + 2f_2  \right]
n' + f_2 m n''
\end{equation}
Using that, to leading order in $\gamma$, $f_2 = \sigma^2/2$, gives us
Eq. (\ref{MZeq}).
\section{The Number of Red Families\label{families}}
The number of red families is, from Eq. (\ref{rf})
\begin{eqnarray}
 F^R&=&\int_0^\infty dx\,\left(1 - e^{-pR_o/N_o}\right)\times \nonumber\\
 &\ &\quad\quad\frac{A}{x} U\left(1+\nu,0,\frac{2\gamma}{\sigma^2(1+\nu)}x\right) 
 \end{eqnarray}
 We cannot break up the term in the parenthesis as is, since both integrals would then diverge.  We therefore regularize the problem:
 \begin{eqnarray}
 F^R &=& A \lim_{\epsilon\to 0} \int_0^\infty\left(1 - e^{-s}\right) x^{\epsilon-1} U\left(1+\nu,0,x\right) dx\nonumber\\
 &=& A \lim_{\epsilon\to 0} \frac{\Gamma(\epsilon)\Gamma(\epsilon+1)}{\Gamma\left(2+\nu+\epsilon\right)} \times\nonumber\\
 &\ &\Big[ {}_2F_1\left(\epsilon,\epsilon +1;2+\nu+\epsilon;1\right) \nonumber\\
 &\ &\quad- {}_2F_1\left(\epsilon,\epsilon +1;2+\nu+\epsilon+1;1-s\right)\Big] \nonumber\\
&=& A \lim_{\epsilon\to 0} \sum_{n=1}^\infty \frac{\Gamma(\epsilon+n)\Gamma(\epsilon+n+1)}{n!\Gamma\left(2+\nu + \epsilon + n \right)} \left[1 - (1-s)^n\right]\nonumber \\
&=& A \sum_{n=0}^\infty  \frac{\Gamma(n+1)\Gamma(n+2)}{(n+1)!\Gamma\left(3+\nu  + n \right)} \left[1 - (1-s)^{n+1}\right] \nonumber\\
&=& \frac{A}{\Gamma\left(3+\nu\right)} \Big[{}_2F_1\left(1,1;3+\nu;1\right)\nonumber\\
&\ &\qquad\qquad{}-(1-s){}_2F_1\left(1,1;3+\nu;1\right)\Big]\nonumber\\
&=& \frac{n_0 \nu}{(2+\nu)s}
\Big[{}_2F_1\left(1,1;3+\nu;1\right)\nonumber\\
&\ &\qquad{}-(1-s)\,{}_2F_1\left(1,1;3+\nu;1-s\right)\Big]
\end{eqnarray}











`




\end{document}